\documentclass[sigconf]{acmart}

\renewcommand\footnotetextcopyrightpermission[1]{} % removes footnote with conference information in first column
\usepackage{trimclip}
\usepackage{pgf}
\newcommand{\clip}[1]{\clipbox{0.2cm 0.2cm 0.44cm 0.18cm}{#1}}

\usepackage{dblfloatfix}
\usepackage{cleveref}

\DeclareMathOperator*{\argmin}{arg\,min}

\setcopyright{none}
\def\BibTeX{{\rm B\kern-.05em{\sc i\kern-.025em b}\kern-.08emT\kern-.1667em\lower.7ex\hbox{E}\kern-.125emX}}

\copyrightyear{2019}
\acmYear{}
\acmConference[]{}{}{}
\acmPrice{}
\acmDOI{}
\acmISBN{}

\begin{document}

%
% The "title" command has an optional parameter, allowing the author to define a "short title" to be used in page headers.
\title{Hamming Sentence Embeddings for Information Retrieval}

% The "author" command and its associated commands are used to define
% the authors and their affiliations.  Of note is the shared
% affiliation of the first two authors, and the "authornote" and
% "authornotemark" commands used to denote shared contribution to the
% research.

% \author{\color{white}-}
% \email{-}
% \affiliation{%
%   \institution{-}
%   \streetaddress{-}
%   \city{-}
%   \postcode{-}
%   \country{-}
% }

\author{Felix Hamann}
\email{felix.hamann@hs-rm.de}
\affiliation{%
  \institution{RheinMain University of Applied Sciences}
  \streetaddress{Unter den Eichen 5}
  \city{Wiesbaden}
  \postcode{65195}
  \country{Germany}
}

\author{Nadja Kurz}
\email{nadja.b.kurz@student.hs-rm.de}
\affiliation{%
  \institution{RheinMain University of Applied Sciences}
  \streetaddress{Unter den Eichen 5}
  \city{Wiesbaden}
  \postcode{65195}
  \country{Germany}
}

\author{Adrian Ulges}
\email{adrian.ulges@hs-rm.de}
\affiliation{%
  \institution{RheinMain University of Applied Sciences}
  \streetaddress{Unter den Eichen 5}
  \city{Wiesbaden}
  \postcode{65195}
  \country{Germany}
}

%
% By default, the full list of authors will be used in the page
% headers. Often, this list is too long, and will overlap other
% information printed in the page headers. This command allows the
% author to define a more concise list of authors' names for this
% purpose.

% - redacted -
\renewcommand{\shortauthors}{Hamann, Kurz and Ulges}

\begin{abstract}
In retrieval applications, binary hashes are known to offer significant improvements in terms of both memory and speed. We investigate the compression of sentence embeddings using a neural encoder-decoder architecture, which is trained by minimizing reconstruction error. Instead of employing the original real-valued embeddings, we use latent representations in Hamming space produced by the encoder for similarity calculations.

In quantitative experiments on several benchmarks for semantic similarity tasks, we show that our compressed hamming embeddings yield a comparable performance  to uncompressed embeddings  (Sent2Vec, InferSent, Glove-BoW), at compression ratios of up to 256:1. We further demonstrate that our model strongly decorrelates input features, and that the compressor generalizes well when pre-trained on Wikipedia sentences. We publish the source code on Github\footnote{https://github.com/ungol-nlp} and all experimental results\footnote{http://bit.ly/2H8cP7Q}.
\end{abstract}

% FIXME http://dl.acm.org/ccs.cfm

%
% Keywords. The author(s) should pick words that accurately describe the work being
% presented. Separate the keywords with commas.

%
% A "teaser" image appears between the author and affiliation information and the body
% of the document, and typically spans the page.
%   \Description{Enjoying the baseball game from the third-base seats. Ichiro Suzuki preparing to bat.}

%
% This command processes the author and affiliation and title information and builds
% the first part of the formatted document.
\maketitle
\thispagestyle{empty}

\section{Introduction}

% word embeddings

Dense, real-valued embeddings play a fundamental role as representations for words~\cite{mikolov2013efficient, pennington2014glove, bojanowski2016enriching} or sentences and documents~\cite{pagliardini2017unsupervised, conneau2017supervised, arora2016simple} 
%for textual entities 
in neural NLP models.
%~\cite{goldberg2016primer}. 
They are known to convey semantics -- for example, the semantic similarity of two embedded entities is commonly expressed by their embeddings' distance.
%Recently,  multiple approaches have been suggested to learn embeddings 

Since embeddings are high-dimensional vectors (often of 100-1000  features), they require memory of up to multiple GB of memory for large vocabularies or corpora. To allow low-capacity devices such as mobile phones or embedded systems to facilitate machine learning models that process word embeddings, recent work implements encoder-decoder models for compressing word embeddings~\cite{tissier2018near, shu2017compressing}, which are trained to reconstruct the original embeddings. The resulting (binary) latent representation of the encoder can then be used for downstream tasks like classification or retrieval, or can be used in conjunction with an instance of the decoder 
%on the target device to create 
to create a reproduction of the original embedding. 
%downstream auf rekonstruktionen
%auf sentiment.
%AAAI:
%- sematic similarity 
%- retrieval
%- nur wort-ebene.
This approach not only reduces the required memory significantly, but also speeds up the distance calculation (either in a brute-force comparison or in combination with index structures~\cite{norouzi2014fast}).

While the above approaches have focused on word-level compression~\cite{tissier2018near, shu2017compressing}, 
 we study the compression of sentence-level embeddings.
 %transfer the problem over to the embedding of sentences. 
This is of practical relevance, since unlike words -- which come with a fixed vocabulary, such that similarities or clusters can be cached -- the space of sentences is virtually infinite.
We apply an encoder-decoder model similar to Shu's and Nakayama's~\cite{shu2017compressing} to several state-of-the-art sentence embeddings (Sent2Vec, InferSent, GloVe-BoW) and use the resulting binary embeddings in textual similarity and retrieval tasks. Our contributions are:

\begin{itemize}
\item We show that the spatial properties of sentence embeddings are retained well by the produced hash codes, which holds regardless of the upstream sentence embedding model. Thereby, accuracy  depends on the task at hand: While 
compressed embeddings yield competitive results for semantic similarity (STS 2012-16), they are outperformed in topic-oriented categorization tasks.
\item We show that a compressor trained on relatively few (100K) Wikipedia sentences in a few minutes generalizes well. %across all evaluated tasks.
\item An explanation for the good performance of our model is that it decorrelates redundant dimensions in the input data, as we demonstrate quantitatively.
\end{itemize}

\section{Related Work}

One of the earliest methods to produce hash codes based on neural architectures was {\it semantic hashing}~\cite{salakhutdinov2009semantic}, which produces bit-codes by training a stack of RBMs learning a latent representation of word frequency. 
%The model is fine-tuned by unfolding the RBMs such that they form an auto-encoder.
In recent work either the combination of existing word embeddings~\cite{wieting2015towards, arora2016simple} or independent machine learning models~\cite{logeswaran2018efficient, conneau2017supervised, le2014distributed, pagliardini2017unsupervised, zhu2015aligning}  transform phrases, sentences or documents to embeddings. To our knowledge, no attempt to create hash codes for such sentence embeddings has been evaluated so far.

On word level, hashing has been studied by introducing a virtual quantization function to the CBOW approach~\cite{lam2018word2bits}. \Citet{shu2017compressing} introduce an auto-encoder for compressing word embeddings but did not evaluate a configuration which produces binary codes. Recently~\cite{tissier2018near} proposed a neural auto-encoder that produces hash codes by thresholding the latent representation. To preserve the spatial information of the input space, a regularization method is added to the loss function.

\section{Approach}

\begin{figure}
  \includegraphics[width=\columnwidth]{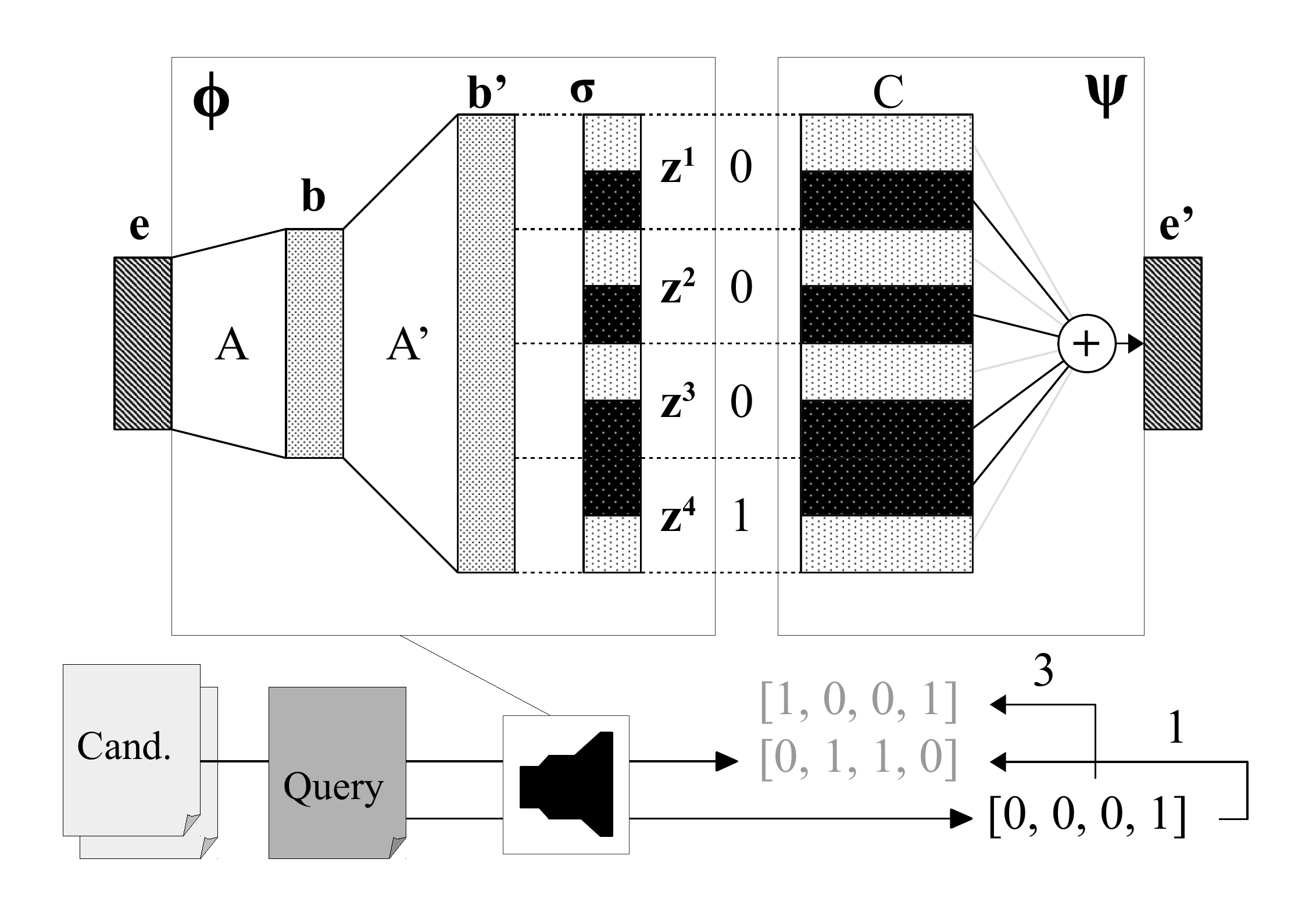}
  \caption{Illustration of our approach: Embeddings ($\mathbf{e}$) are reconstructed ($\mathbf{e}'$) by an encoder-($\phi$)decoder($\Psi$). The resulting binary bottleneck representations $\mathbf{z}_i$ are used for a low-resource comparison in retrieval (bottom). \vspace{-1em} \label{fig:sts_models}}
\end{figure}

We explore the usefulness of hashes for fast retrieval using a model similar to Shu's and Nakayama's \cite{shu2017compressing}.
Figure \ref{fig:sts_models} illustrates our approach: Sentence embeddings (gained by a pre-trained model) are compressed using an encoder-decoder architecture (in the following referred to as the {\it compressor}) which is trained to minimize reconstruction error. The compressor consists of an encoder \( \phi \) and a decoder \( \psi \).
The encoder's output is mapped to a binary vector which is then used for a low-resource comparison in retrieval.
%\subsection{Auto-Encoder}

\quad \\ 

{\bf The encoder} transforms a real-valued embedding $\mathbf{e}$ 
%of dimensionality \( d \) 
to $b$ one-hot encoded representations, each of the same dimension $p$. 
% As we found $p=2$ to work best,
We use a fixed value of $p=2$ (which we found to work best) in this paper, such that the bottleneck is  a $(2\cdot b)$-dimensional binary feature. More precisely, 
the encoder consists of two fully connected layers: $\phi(\mathbf{e}) := \sigma(f_2(f_1(\mathbf{e})))$ where
%\( f_1 \):
%and \( f_2 \) and a special function \( \sigma \) which controls the one-hot encoding of the logits offered by the neural activation of these layers (assuming all input vectors are encoded as column vectors). The encoder \( \phi \) is defined as:
% inline formula
% \( \phi_{\Theta}(\mathbf{e}) &:= \sigma(f_2 \circ f_1)(\mathbf{e}) \)
% with
% \( f_1 (\mathbf{x}) := \text{tanh}(A \cdot \mathbf{x} + \mathbf{b}) \)
% and
% \( f_2 (\mathbf{x}) := \text{softplus}(A' \cdot \mathbf{x} +
% \mathbf{b}' \).
% multi-line formula
\begin{align}\label{eq:compr-enc}
  \nonumber
   f_1 (\mathbf{x}) := \text{tanh}(A \cdot \mathbf{x} + \mathbf{b}) \text{\quad} & \text{and \quad} f_2 (\mathbf{x}) := \text{softplus}(A' \cdot \mathbf{x} + \mathbf{b}')
\end{align}

with $A \in \mathbb{R}^{b \times d}$ and $A' \in \mathbb{R}^{2b \times b}$.  
%The number of output neurons of the second layer \( f_2 \) is \( 2b \) by definition. 
The first layer \( f_1 \) has \( b \) output neurons, the second layer produces $2b$ logit values $(y_1^1,y_2^1), (y_1^2, y_2^2),$  $\dots,(y_1^b, y_2^b)$ viewed in pairs, $\mathbf{y}^k := (y_1^k, y_2^k)$. The function \( \sigma \) applies a  softmax to each pair, i.e.
$
 \sigma( \mathbf{y}^1, 
 ...,
 \mathbf{y}^b) = 
 \mathbf{z}^1, 
 ...,
 \mathbf{z}^b
$ with
\begin{equation*}
  z_i^k := \frac{\exp(l(y_i^k))}{\exp(l(y_1^k)) + \exp(l(y_2^k))}
  \; \text{with} \;
  l(y) = \frac{1}{\tau} \cdot (y + g).
\end{equation*}
With the function $l(\cdot)$, we apply the Gumbel Softmax trick~\cite{jang2016categorical, maddison2016concrete} to bias the model towards a categorical distribution, whereas the free parameter \( \tau \) controls the \textit{degree} of discreteness and $g$ denotes random noise sampled from the Gumbel extreme value distribution \cite{gumbel1954statistical}. When applying the trained model, we set $g=0$ and obtain a $b$-dimensional binary representation by thresholding $(z_1^1,z_1^2,...,z_1^b)$ at 0.5. This representation is referred to as the {\it compressed} embedding in the following, and is used for similarity evaluation.

%Values below one enforce a higher tendency towards discrete distributions. The function \( \sigma \) extends the softmax function such that iid noise terms \( g_k \sim G \) are added to each output activation and weighted by the annealing parameter \( \tau \):
%controls the one-hot encoding of the encoder's output activation for each code component. This means the function operates on \( b \) pairs \( \mathbf{y}' := (\mathbf{y}_i, \mathbf{y}_{i+1}), 2|i  \). 

%only as binary codes are considered and thus the possible values for each component are either zero or one (\( n = 2 \)). 
\quad

{\bf The decoder} transforms the bottleneck back to a reconstruction $\mathbf{e}'$ of the input embedding $\mathbf{e}$ %reconstructed float embedding vector of size \( d \). Given an embedding vector \( \mathbf{e} \), the model produces a reconstruction \( \mathbf{e}' \) 
%by computing \( \mathbf{e}' := (\psi_{\Theta'} \circ \phi_{\Theta})(\mathbf{e})\).
. It
is based on the idea of additive vector quantization for compression \cite{babenko2014additive, jegou2011product}, i.e. the reconstruction is a linear combination of basis vectors
organised in codebooks. There are $2b$ such basis vectors stored in a matrix \( C \in \mathbb{R}^{d \times 2b} \).
%Each component of the code is now assigned two basis vectors (the codebook of the component) 
%and the reconstruction is the weighted sum over these vectors . 
Given an encoder output \( \mathbf{x} = (\mathbf{z}^1, ..., \mathbf{z}^b)^T \) of size \( 2b \), the decoder output \( \mathbf{e}' \) is defined as \( \psi(\mathbf{x}) := C \cdot \mathbf{x} \).

\quad

\textbf{Training:} Given a set of training embeddings $\mathbf{e}_1,\mathbf{e}_2,...,\mathbf{e}_n$, the model's parameters \( \{A, \mathbf{b}, A', \mathbf{b}', C\} \) are fitted such that the average Euclidean distance of all training samples to their respective reconstruction is minimized:
\begin{equation*}
  \argmin_{A, \mathbf{b}, A', \mathbf{b}', C}
  \;\;\;\;
  \sum_{i=1}^n ||\mathbf{e}_{i} - \mathbf{e}'_{i}||_{_2}^{^2}
\end{equation*}
Optimization is carried out using Stochastic Gradient Descent with the Adam Optimizer (initial learning rate \(10^{-4}\)). 
%The trainable parameters of the network are
%\( \{A, \mathbf{b}, A', \mathbf{b}', C\} \). 
The decoder's codebook's vectors $c_i$ are initialised by randomly sampling from the training embeddings \( \mathbf{e}_i \): \( c_i = \frac{1}{xb} \mathbf{e}_i \) with \( x \in [1,2] \) to obtain about the same norm for the reconstruction when training starts. Training is stopped when no significant change ($ \Delta < 10^{-5} $ to $ 10^{-4} $) of the loss value can be observed over a fixed period of $100$ epochs. The parameter $ \tau $ is set to one for the majority of models\footnote{We found that very few models only trained by annealing from $ \tau = 1$ to $ 0.75 $.}.

\section{Experiments}

\begin{figure*}
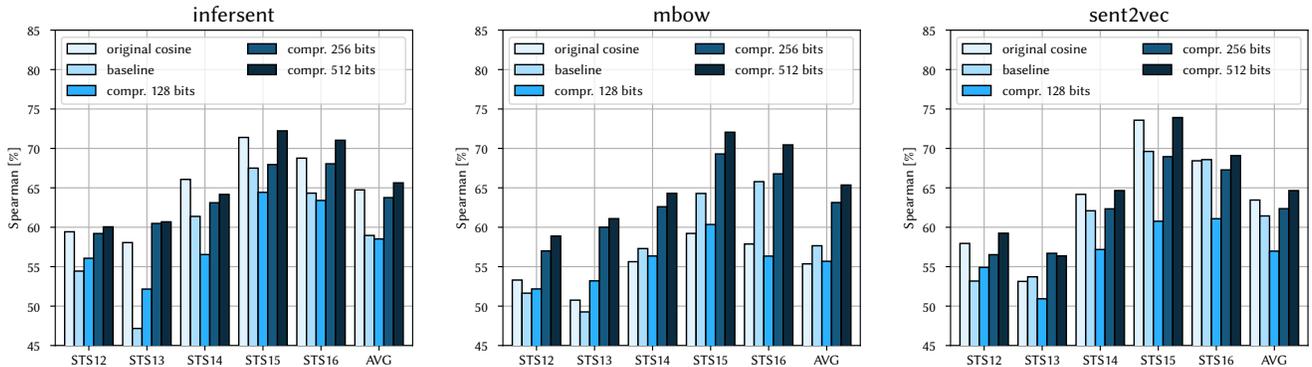

    \begin{minipage}[b]{0.33\textwidth}
        \centering
        \resizebox{\linewidth}{!}{\clip{
            \input{sts-infersent-comparison.pgf}
          }}
    \end{minipage}
    \begin{minipage}[b]{0.33\textwidth}
        \centering
        \resizebox{\linewidth}{!}{\clip{
            \input{sts-mbow-comparison.pgf}
          }}
    \end{minipage}
    \begin{minipage}[b]{0.33\textwidth}
        \centering
        \resizebox{\linewidth}{!}{\clip{
            \input{sts-sent2vec-comparison.pgf}
          }}
    \end{minipage}
    \caption{The correlation between ground truth
 and computed similarities for STS12 to STS16 (higher is better).}
    \label{fig:sts-bits}
\end{figure*}

To examine how well the spatial information of sentence embeddings is retained when compressed with our encoder,
we use two similarity-based tasks, namely semantic textural similarity (STS, Section \ref{sts}) and k-NN document classification (Section \ref{knn}). We compare three types of embeddings:
\begin{itemize}
    \item The {\bf original} real-valued sentence embeddings, using InferSent~\cite{conneau2017supervised}, Sent2Vec~\cite{pagliardini2017unsupervised}, or bag-of-words with averaged Glove vectors~\cite{pennington2014glove}. All models were used in their pre-trained versions, no fine-tuning was applied. We tested several common distance measures but report the cosine similarity (which we found to work best).  %unless stated otherwise.
    \item The {\bf compressed} sentence embeddings from our encoder. We trained compressors for $128$, $256$ and $512$ bits, either %on a number of 
    randomly sampled sentences from Wikipedia 
    %varying between $10,000$ and $10$ mio., 
    or on sentences collected from the respective task's training set (e.g., $10518$ sentences from the STS datasets). Any binary embeddings are compared using the Hamming distance.
    \item a simple \textbf{baseline} binarization, which transforms a real-valued input embedding $\mathbf{e} \in \mathbb{R}^d$ into a binary embedding $\mathbf{e}' \in \{0, 1\}^d$ by thresholding each dimension at the median:
$    \mathbf{e}'_{i} := \mathbf{1}_{\mathbf{e}_i \geq \, \text{median}(E_{i,*})}
$
(where $
E = [\mathbf{e}_1,
     ...,
     \mathbf{e}_n] \in \mathbb{R}^{d \times n}
$ denotes the embedding matrix).
\end{itemize}

%Another interesting question is how much training data our compressor requires, and how well it generalizes from one text domain to another. To investigate this question, :

\subsection{Semantic Similarity}
\label{sts}

%We evaluate a set of different model configurations focusing on the impact of both bit-size of the produced codes and the selection of training samples for the auto-encoder. The hash codes are compared by their respective hamming distance in a series of semantic textual similarity tasks to measure to which degree the produced codes still encode semantic similarity by distance.

We use the SentEval
\cite{conneau2018senteval} implementation of the SemEval Semantic Textual Similarity (STS) tasks 2012-2016. \Cref{fig:sts-bits} compares our compressor model (trained on 1 mio Wikipedia sentences) with the other methods. We report Spearman's $\rho$ between ground truth similarity annotations and similarity scores (higher is better).

Our model achieves competitive results for all three sentence embeddings, even outperforming the original real-valued embeddings. This improvement is strongest for averaged word vectors ( \textit{mbow}). Also, Figure \ref{fig:sts_models} shows that our model generalizes well from Wikipedia to other text domains without the need of any refinement, and that training on a relatively small dataset (100,000 sentences) -- which takes in the order of minutes -- already yields competitive performance. %Training a 512-bit compressor on this dataset takes in the order of minutes. 
%Training on 1 mio. sentences gives small improvements, training on 10 mio. sentences does not seem necessary. Therefore, 
For the remainder of the paper, models trained on 1 mio. Wikipedia sentences will be used.

\begin{figure}
    \resizebox{\linewidth}{!}{\clip{
        \input{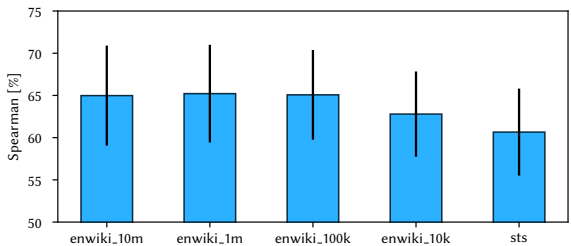}
    }}
    \caption{Training the compressor on Wikipedia outperforms training on the target domain for STS. \vspace{-1em}}
    \label{fig:sts-enwiki}
\end{figure}

\begin{figure*}
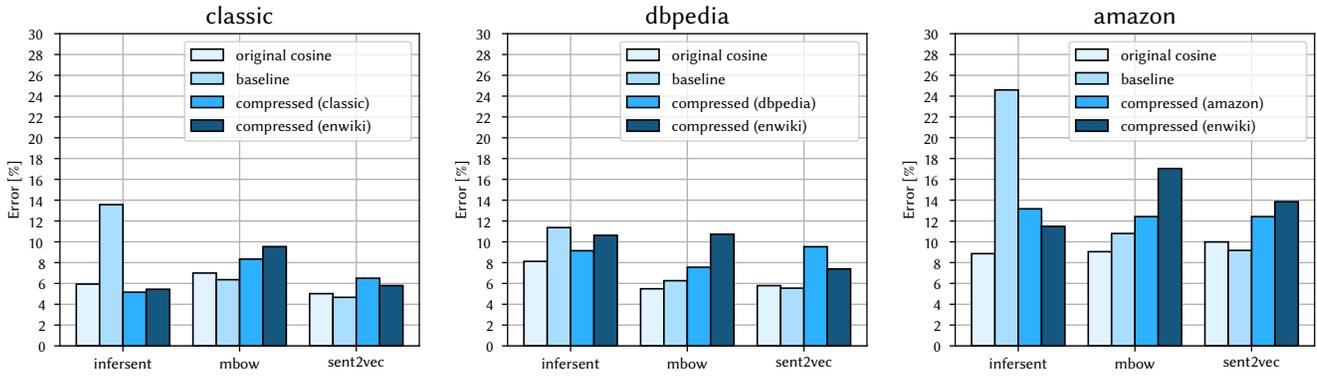

    \begin{minipage}[b]{0.33\textwidth}
        \centering
        \resizebox{\linewidth}{!}{\clip{
            \input{ir-classic-comparison.pgf}
        }}
    \end{minipage}
    \begin{minipage}[b]{0.33\textwidth}
        \centering
        \resizebox{\linewidth}{!}{\clip{
            \input{ir-dbpedia-comparison.pgf}
        }}
    \end{minipage}
    \begin{minipage}[b]{0.33\textwidth}
        \centering
        \resizebox{\linewidth}{!}{\clip{
            \input{ir-amazon-comparison.pgf}
        }}
    \end{minipage}
    \caption{Error for the three upstream sentence embedding methods per retrieval dataset.\vspace{-1em}}
    \label{fig:ir-error}
\end{figure*}

\subsection{Memory Footprint}

The biggest advantage of the binary representation is the compactness of its encoding. For example, storing 10 mio. 700-dimensional \textit{sent2vec} embeddings takes 28GB of disk space ($700 \times 4$ bytes per embedding at single float precision), while the corresponding (512-bit) hash codes require only $640$ MB while reaching the same accuracy. For Infersent ($4096$ dimensions), memory is even reduced from $163$GB to $640$MB (which corresponds to a reduction factor of 256:1). % while reaching the same accuracy.

\subsection{Correlation Inspection}
The competitive performance of our compressor could be due to the fact that the
original float embeddings contain redundant features (i.e. certain dimensions in the embeddings carry the same information), which may have an impact on semantic similarity calculations. In contrast to this, the compressor -- forced to remove redundancy in the embeddings -- aggregates these dimensions.  
%Note that -- while other binarization models are specifically trained to remove redundant information~\cite{tissier2018near} -- this is implicit in our compressor.

To test this hypothesis, we inspect the scale of correlations between dimensions in the embeddings: Let $\mathbf{e}_1,...,\mathbf{e}_n$ denote d-dimensional embeddings, and let $\rho(i,j)$ (for $i,j = 1,...,d$) denote the correlation between two dimensions in these embeddings. We measure the overall correlation within the embeddings as
\begin{equation}
\frac{1}{d^2} \sum_{i=1}^d \sum_{j=1}^d | \rho (i,j) |.
\label{eq:correlation}
\end{equation}
Table \ref{tab:correlation} illustrates the overall correlation on the Wikipedia-1m dataset, using (a) the original real-valued  embeddings, and (b) the 512-dimensional hamming embeddings. We observe a strong reduction of correlation between the embeddings by a factor of about 3-4, which illustrates the decorrelating effect of our compressor.

\begin{table}[bp]
\centering
\begin{tabular}{l||r|r}
& \multicolumn{2}{c}{\bf avg. correlation (\%)} \\
embedding type & original & compressed \\
\hline
\hline
infersent & 10.17 &  2.26 \\
\hline
mbow      &  9.31 &  2.73 \\
\hline
sent2vec  &  5.37 &  1.83 \\
\end{tabular}
\caption{The average absolute correlation of sentence embedding's dimensions (Equation \eqref{eq:correlation}) on the Wikipedia-1mio dataset. The compressed embeddings' correlation is about 3-4 lower than the original embeddings'. \vspace{-1em}}
\label{tab:correlation}
\end{table}

% \begin{table}[H]
% \centering
% \resizebox{\linewidth}{!}{\begin{tabular}{l||rrrrrr}
% reference embedding & dim & cosine & euclidean & manhattan & baseline & enwiki\_1m 512\\
% \hline
% \hline
% infersent & 4096 & 64.7(\pm 5.8) & 62.4 (\pm 5.0) & 57.7 (\pm 5.2) & 59.0 (\pm 8.2) & 65.6 (\pm 5.7) \\
% mbow & 300 & 55.4 (\pm 3.4) & 55.6 (\pm 4.1) & 56.1 (\pm 4.2) & 57.7 (\pm 7.4) & 65.4 (\pm 5.7) \\
% sent2vec & 700 & 63.5 (\pm 8.1) & 52.3 (\pm 5.4) & 52.3 (\pm 5.3) & 61.4 (\pm 7.9) & 64.7 (\pm 7.1) \\

% \end{tabular}}
% \caption{The average spearman rank correlation coefficient and standard deviation on the STS'12-16 datasets for different similarity measures. FIXME}
% \label{tab:}
% \end{table}

\subsection{k-NN Document Classification}
\label{knn}
Finally, we assess how well the Hamming sentence embeddings can be used in a more heterogeneous, topic-oriented information retrieval setting. We employ three datasets: (1)  \textbf{amazon}~\cite{dataset:amazon07} (reviews from four different product categories; 6400/1600 documents), (2) \textbf{classic}~\cite{dataset:classic} (short citations from academic papers of four different categories; 5600/1400 documents) and (3) \textbf{dbpedia}~(self-crawled Wikipedia pages sampled from 13 classes; 47000/11400 documents; inspired by~\cite{dataset:zhang15}). All documents are pruned to the first 1000 tokens. Each test document is classified by a voting over its $10$ nearest training documents (computed by their similarity scores on our different embeddings), whereas
each vote is weighted such that for the \( n \)-th neighbour the weight for the vote is defined as \( 1 / \sqrt{n} \).

\begin{figure}
  \includegraphics[width=\columnwidth]{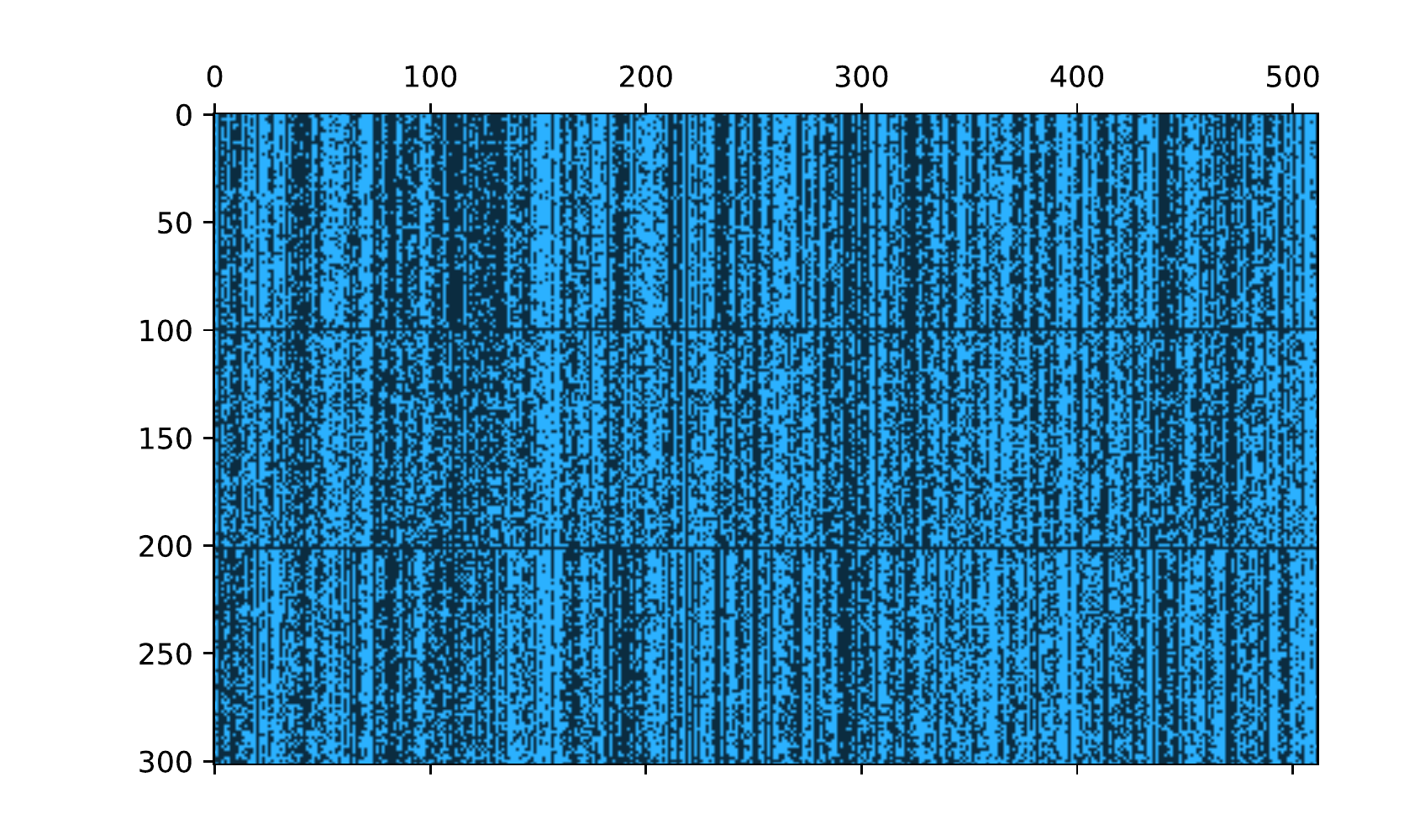}
  \caption{Random binary embeddings from the dbpedia classes \textit{Film} (top), \textit{Athlete} (center) and \textit{Plant} (bottom).% of the \textit{dbpedia} data set. 
  \vspace{-1em} \label{fig:bits}}
\end{figure}

The results of this evaluation are mixed (\Cref{fig:ir-error}): Working with the original embeddings and determining the cosine similarity works best in general. The median baseline works surprisingly well, in some cases even outperforming all other methods, but failing for Infersent vectors on {\it classic} and {\it amazon}. 
%Except for \textbf{infersent} the simple binarization always outperforms the compressor trained on the dataset's training data and the \textbf{enwiki} compressor. {\color{red} Pretty short, more?}
Also, the picture with respect to the training domain is mixed, as sometimes training on the target domain and sometimes training on Wikipedia performs better. Overall, we found results for this topical evaluation to be more instable than for sentence similarity. Figure \ref{fig:bits} matches these quantitative results with a visualization of embeddings: For the classes {\it Film} (top), {\it Athlete} (center) and {\it Plant} (bottom) from the dbpedia dataset, we picked 300 random samples and visualized the binary hash codes ($b=512$, trained on $1$ mio. Wikipedia sentences). The figure illustrates that the binary hashes show some, but not a strong correlation to the document category, an issue that will require further investigation.

\section{Conclusion}

In this paper, we have studied a neural encoder-decoder to produce binary hash codes for an efficient similarity matching on sentence- and paragraph level. We have shown that the spatial information is retained well on a simple sentence-level similarity task. Also, we found our model to decorrelate the input embeddings' dimensions, 
and training on a limited number of Wikipedia sentences generalizes well (at least for the STS task). When applied to topic-oriented k-NN classification task, the model yields mixed results. We publish a spreadsheet with all experimental results and the source code.

% The acknowledgments section is defined using the "acks" environment
% (and NOT an unnumbered section). This ensures the proper
% identification of the section in the article metadata, and the
% consistent spelling of the heading.

%\begin{acks}
%{\color{red} -redacted-}
%\end{acks}

% The next two lines define the bibliography style to be used, and the
% bibliography file.
\bibliographystyle{ACM-Reference-Format}
% \bibliography{references}


\begin{thebibliography}{24}

%%% the Web address.

\ifx \showCODEN    \undefined \def \showCODEN     #1{\unskip}     \fi
\ifx \showDOI      \undefined \def \showDOI       #1{#1}\fi
\ifx \showISBNx    \undefined \def \showISBNx     #1{\unskip}     \fi
\ifx \showISBNxiii \undefined \def \showISBNxiii  #1{\unskip}     \fi
\ifx \showISSN     \undefined \def \showISSN      #1{\unskip}     \fi
\ifx \showLCCN     \undefined \def \showLCCN      #1{\unskip}     \fi
\ifx \shownote     \undefined \def \shownote      #1{#1}          \fi
\ifx \showarticletitle \undefined \def \showarticletitle #1{#1}   \fi
\ifx \showURL      \undefined \def \showURL       {\relax}        \fi
% invisible to TeX
\providecommand\bibfield[2]{#2}
\providecommand\bibinfo[2]{#2}
\providecommand\natexlab[1]{#1}
\providecommand\showeprint[2][]{arXiv:#2}

\bibitem[\protect\citeauthoryear{Arora, Liang, and Ma}{Arora
  et~al\mbox{.}}{2016}]%
        {arora2016simple}
\bibfield{author}{\bibinfo{person}{Sanjeev Arora}, \bibinfo{person}{Yingyu
  Liang}, {and} \bibinfo{person}{Tengyu Ma}.} \bibinfo{year}{2016}\natexlab{}.
\newblock \showarticletitle{A simple but tough-to-beat baseline for sentence
  embeddings}.
\newblock  (\bibinfo{year}{2016}).
\newblock


\bibitem[\protect\citeauthoryear{Babenko and Lempitsky}{Babenko and
  Lempitsky}{2014}]%
        {babenko2014additive}
\bibfield{author}{\bibinfo{person}{Artem Babenko} {and} \bibinfo{person}{Victor
  Lempitsky}.} \bibinfo{year}{2014}\natexlab{}.
\newblock \showarticletitle{Additive quantization for extreme vector
  compression}. In \bibinfo{booktitle}{\emph{Proceedings of the IEEE Conference
  on Computer Vision and Pattern Recognition}}. \bibinfo{pages}{931--938}.
\newblock


\bibitem[\protect\citeauthoryear{Blitzer, Dredze, and Pereira}{Blitzer
  et~al\mbox{.}}{2007}]%
        {dataset:amazon07}
\bibfield{author}{\bibinfo{person}{John Blitzer}, \bibinfo{person}{Mark
  Dredze}, {and} \bibinfo{person}{Fernando Pereira}.}
  \bibinfo{year}{2007}\natexlab{}.
\newblock \showarticletitle{Biographies, bollywood, boom-boxes and blenders:
  Domain adaptation for sentiment classification}. In
  \bibinfo{booktitle}{\emph{ACL '07}}. \bibinfo{pages}{440--447}.
\newblock


\bibitem[\protect\citeauthoryear{Bojanowski, Grave, Joulin, and
  Mikolov}{Bojanowski et~al\mbox{.}}{2016}]%
        {bojanowski2016enriching}
\bibfield{author}{\bibinfo{person}{Piotr Bojanowski}, \bibinfo{person}{Edouard
  Grave}, \bibinfo{person}{Armand Joulin}, {and} \bibinfo{person}{Tomas
  Mikolov}.} \bibinfo{year}{2016}\natexlab{}.
\newblock \showarticletitle{Enriching word vectors with subword information.
  CoRR abs/1607.04606}.
\newblock \bibinfo{journal}{\emph{URL http://arxiv. org/abs/1607.04606}}
  (\bibinfo{year}{2016}).
\newblock


\bibitem[\protect\citeauthoryear{Conneau and Kiela}{Conneau and Kiela}{2018}]%
        {conneau2018senteval}
\bibfield{author}{\bibinfo{person}{Alexis Conneau} {and} \bibinfo{person}{Douwe
  Kiela}.} \bibinfo{year}{2018}\natexlab{}.
\newblock \showarticletitle{SentEval: An Evaluation Toolkit for Universal
  Sentence Representations}.
\newblock \bibinfo{journal}{\emph{arXiv preprint arXiv:1803.05449}}
  (\bibinfo{year}{2018}).
\newblock


\bibitem[\protect\citeauthoryear{Conneau, Kiela, Schwenk, Barrault, and
  Bordes}{Conneau et~al\mbox{.}}{2017}]%
        {conneau2017supervised}
\bibfield{author}{\bibinfo{person}{Alexis Conneau}, \bibinfo{person}{Douwe
  Kiela}, \bibinfo{person}{Holger Schwenk}, \bibinfo{person}{Loic Barrault},
  {and} \bibinfo{person}{Antoine Bordes}.} \bibinfo{year}{2017}\natexlab{}.
\newblock \showarticletitle{Supervised learning of universal sentence
  representations from natural language inference data}.
\newblock \bibinfo{journal}{\emph{arXiv preprint arXiv:1705.02364}}
  (\bibinfo{year}{2017}).
\newblock


\bibitem[\protect\citeauthoryear{Gumbel}{Gumbel}{1954}]%
        {gumbel1954statistical}
\bibfield{author}{\bibinfo{person}{Emil~Julius Gumbel}.}
  \bibinfo{year}{1954}\natexlab{}.
\newblock \showarticletitle{Statistical theory of extreme values and some
  practical applications}.
\newblock \bibinfo{journal}{\emph{NBS Applied Mathematics Series}}
  \bibinfo{volume}{33} (\bibinfo{year}{1954}).
\newblock


\bibitem[\protect\citeauthoryear{Jang, Gu, and Poole}{Jang
  et~al\mbox{.}}{2016}]%
        {jang2016categorical}
\bibfield{author}{\bibinfo{person}{Eric Jang}, \bibinfo{person}{Shixiang Gu},
  {and} \bibinfo{person}{Ben Poole}.} \bibinfo{year}{2016}\natexlab{}.
\newblock \showarticletitle{Categorical reparameterization with
  gumbel-softmax}.
\newblock \bibinfo{journal}{\emph{arXiv preprint arXiv:1611.01144}}
  (\bibinfo{year}{2016}).
\newblock


\bibitem[\protect\citeauthoryear{Jegou, Douze, and Schmid}{Jegou
  et~al\mbox{.}}{2011}]%
        {jegou2011product}
\bibfield{author}{\bibinfo{person}{Herve Jegou}, \bibinfo{person}{Matthijs
  Douze}, {and} \bibinfo{person}{Cordelia Schmid}.}
  \bibinfo{year}{2011}\natexlab{}.
\newblock \showarticletitle{Product quantization for nearest neighbor search}.
\newblock \bibinfo{journal}{\emph{IEEE transactions on pattern analysis and
  machine intelligence}} \bibinfo{volume}{33}, \bibinfo{number}{1}
  (\bibinfo{year}{2011}), \bibinfo{pages}{117--128}.
\newblock


\bibitem[\protect\citeauthoryear{Lam}{Lam}{2018}]%
        {lam2018word2bits}
\bibfield{author}{\bibinfo{person}{Maximilian Lam}.}
  \bibinfo{year}{2018}\natexlab{}.
\newblock \showarticletitle{Word2Bits-Quantized Word Vectors}.
\newblock \bibinfo{journal}{\emph{arXiv preprint arXiv:1803.05651}}
  (\bibinfo{year}{2018}).
\newblock


\bibitem[\protect\citeauthoryear{Le and Mikolov}{Le and Mikolov}{2014}]%
        {le2014distributed}
\bibfield{author}{\bibinfo{person}{Quoc Le} {and} \bibinfo{person}{Tomas
  Mikolov}.} \bibinfo{year}{2014}\natexlab{}.
\newblock \showarticletitle{Distributed representations of sentences and
  documents}. In \bibinfo{booktitle}{\emph{International conference on machine
  learning}}. \bibinfo{pages}{1188--1196}.
\newblock


\bibitem[\protect\citeauthoryear{Logeswaran and Lee}{Logeswaran and
  Lee}{2018}]%
        {logeswaran2018efficient}
\bibfield{author}{\bibinfo{person}{Lajanugen Logeswaran} {and}
  \bibinfo{person}{Honglak Lee}.} \bibinfo{year}{2018}\natexlab{}.
\newblock \showarticletitle{An efficient framework for learning sentence
  representations}.
\newblock \bibinfo{journal}{\emph{arXiv preprint arXiv:1803.02893}}
  (\bibinfo{year}{2018}).
\newblock


\bibitem[\protect\citeauthoryear{Maddison, Mnih, and Teh}{Maddison
  et~al\mbox{.}}{2016}]%
        {maddison2016concrete}
\bibfield{author}{\bibinfo{person}{Chris~J Maddison}, \bibinfo{person}{Andriy
  Mnih}, {and} \bibinfo{person}{Yee~Whye Teh}.}
  \bibinfo{year}{2016}\natexlab{}.
\newblock \showarticletitle{The concrete distribution: A continuous relaxation
  of discrete random variables}.
\newblock \bibinfo{journal}{\emph{arXiv preprint arXiv:1611.00712}}
  (\bibinfo{year}{2016}).
\newblock


\bibitem[\protect\citeauthoryear{Mikolov, Chen, Corrado, and Dean}{Mikolov
  et~al\mbox{.}}{2013}]%
        {mikolov2013efficient}
\bibfield{author}{\bibinfo{person}{Tomas Mikolov}, \bibinfo{person}{Kai Chen},
  \bibinfo{person}{Greg Corrado}, {and} \bibinfo{person}{Jeffrey Dean}.}
  \bibinfo{year}{2013}\natexlab{}.
\newblock \showarticletitle{Efficient estimation of word representations in
  vector space}.
\newblock \bibinfo{journal}{\emph{arXiv preprint arXiv:1301.3781}}
  (\bibinfo{year}{2013}).
\newblock


\bibitem[\protect\citeauthoryear{Norouzi, Punjani, and Fleet}{Norouzi
  et~al\mbox{.}}{2014}]%
        {norouzi2014fast}
\bibfield{author}{\bibinfo{person}{Mohammad Norouzi}, \bibinfo{person}{Ali
  Punjani}, {and} \bibinfo{person}{David~J Fleet}.}
  \bibinfo{year}{2014}\natexlab{}.
\newblock \showarticletitle{Fast exact search in hamming space with multi-index
  hashing}.
\newblock \bibinfo{journal}{\emph{IEEE transactions on pattern analysis and
  machine intelligence}} \bibinfo{volume}{36}, \bibinfo{number}{6}
  (\bibinfo{year}{2014}), \bibinfo{pages}{1107--1119}.
\newblock


\bibitem[\protect\citeauthoryear{Pagliardini, Gupta, and Jaggi}{Pagliardini
  et~al\mbox{.}}{2017}]%
        {pagliardini2017unsupervised}
\bibfield{author}{\bibinfo{person}{Matteo Pagliardini},
  \bibinfo{person}{Prakhar Gupta}, {and} \bibinfo{person}{Martin Jaggi}.}
  \bibinfo{year}{2017}\natexlab{}.
\newblock \showarticletitle{Unsupervised learning of sentence embeddings using
  compositional n-gram features}.
\newblock \bibinfo{journal}{\emph{arXiv preprint arXiv:1703.02507}}
  (\bibinfo{year}{2017}).
\newblock


\bibitem[\protect\citeauthoryear{Pennington, Socher, and Manning}{Pennington
  et~al\mbox{.}}{2014}]%
        {pennington2014glove}
\bibfield{author}{\bibinfo{person}{Jeffrey Pennington},
  \bibinfo{person}{Richard Socher}, {and} \bibinfo{person}{Christopher
  Manning}.} \bibinfo{year}{2014}\natexlab{}.
\newblock \showarticletitle{Glove: Global vectors for word representation}. In
  \bibinfo{booktitle}{\emph{Proceedings of the 2014 conference on empirical
  methods in natural language processing (EMNLP)}}.
  \bibinfo{pages}{1532--1543}.
\newblock


\bibitem[\protect\citeauthoryear{Salakhutdinov and Hinton}{Salakhutdinov and
  Hinton}{2009}]%
        {salakhutdinov2009semantic}
\bibfield{author}{\bibinfo{person}{Ruslan Salakhutdinov} {and}
  \bibinfo{person}{Geoffrey Hinton}.} \bibinfo{year}{2009}\natexlab{}.
\newblock \showarticletitle{Semantic hashing}.
\newblock \bibinfo{journal}{\emph{International Journal of Approximate
  Reasoning}} \bibinfo{volume}{50}, \bibinfo{number}{7} (\bibinfo{year}{2009}),
  \bibinfo{pages}{969--978}.
\newblock


\bibitem[\protect\citeauthoryear{Shu and Nakayama}{Shu and Nakayama}{2017}]%
        {shu2017compressing}
\bibfield{author}{\bibinfo{person}{Raphael Shu} {and} \bibinfo{person}{Hideki
  Nakayama}.} \bibinfo{year}{2017}\natexlab{}.
\newblock \showarticletitle{Compressing Word Embeddings via Deep Compositional
  Code Learning}.
\newblock \bibinfo{journal}{\emph{arXiv preprint arXiv:1711.01068}}
  (\bibinfo{year}{2017}).
\newblock


\bibitem[\protect\citeauthoryear{Tissier, Habrard, and Gravier}{Tissier
  et~al\mbox{.}}{2018}]%
        {tissier2018near}
\bibfield{author}{\bibinfo{person}{Julien Tissier}, \bibinfo{person}{Amaury
  Habrard}, {and} \bibinfo{person}{Christophe Gravier}.}
  \bibinfo{year}{2018}\natexlab{}.
\newblock \showarticletitle{Near-lossless Binarization of Word Embeddings}.
\newblock \bibinfo{journal}{\emph{arXiv preprint arXiv:1803.09065}}
  (\bibinfo{year}{2018}).
\newblock


\bibitem[\protect\citeauthoryear{Tunali}{Tunali}{2010}]%
        {dataset:classic}
\bibfield{author}{\bibinfo{person}{Volkan Tunali}.}
  \bibinfo{year}{2010}\natexlab{}.
\newblock \bibinfo{title}{{Data Mining Research, Classic3 and Classic4
  DataSets}}.
\newblock
\newblock
\urldef\tempurl%
\url{http://www.dataminingresearch.com/index.php/2010/09/classic3-classic4-datasets/}
\showURL{%
Retrieved 2019-01-14 from \tempurl}


\bibitem[\protect\citeauthoryear{Wieting, Bansal, Gimpel, and Livescu}{Wieting
  et~al\mbox{.}}{2015}]%
        {wieting2015towards}
\bibfield{author}{\bibinfo{person}{John Wieting}, \bibinfo{person}{Mohit
  Bansal}, \bibinfo{person}{Kevin Gimpel}, {and} \bibinfo{person}{Karen
  Livescu}.} \bibinfo{year}{2015}\natexlab{}.
\newblock \showarticletitle{Towards universal paraphrastic sentence
  embeddings}.
\newblock \bibinfo{journal}{\emph{arXiv preprint arXiv:1511.08198}}
  (\bibinfo{year}{2015}).
\newblock


\bibitem[\protect\citeauthoryear{Zhang and LeCun}{Zhang and LeCun}{2015}]%
        {dataset:zhang15}
\bibfield{author}{\bibinfo{person}{Xiang Zhang} {and} \bibinfo{person}{Yann
  LeCun}.} \bibinfo{year}{2015}\natexlab{}.
\newblock \showarticletitle{Text Understanding from Scratch}.
\newblock \bibinfo{journal}{\emph{CoRR}}  \bibinfo{volume}{abs/1502.01710}
  (\bibinfo{year}{2015}).
\newblock
\showeprint[arxiv]{1502.01710}
\urldef\tempurl%
\url{http://arxiv.org/abs/1502.01710}
\showURL{%
\tempurl}


\bibitem[\protect\citeauthoryear{Zhu, Kiros, Zemel, Salakhutdinov, Urtasun,
  Torralba, and Fidler}{Zhu et~al\mbox{.}}{2015}]%
        {zhu2015aligning}
\bibfield{author}{\bibinfo{person}{Yukun Zhu}, \bibinfo{person}{Ryan Kiros},
  \bibinfo{person}{Rich Zemel}, \bibinfo{person}{Ruslan Salakhutdinov},
  \bibinfo{person}{Raquel Urtasun}, \bibinfo{person}{Antonio Torralba}, {and}
  \bibinfo{person}{Sanja Fidler}.} \bibinfo{year}{2015}\natexlab{}.
\newblock \showarticletitle{Aligning books and movies: Towards story-like
  visual explanations by watching movies and reading books}. In
  \bibinfo{booktitle}{\emph{Proceedings of the IEEE international conference on
  computer vision}}. \bibinfo{pages}{19--27}.
\newblock


\end{thebibliography}

%%% Do NOT edit. File created by style

% If your work has an appendix, this is the place to put it.
% \appendix

\end{document}